\documentclass[superscriptaddress,twocolumn,floatfix,aps,reprint,preprintnumbers]{revtex4-1}

\usepackage{soul}
\usepackage{textcomp}
\usepackage{graphicx}
\usepackage{amsmath}
\usepackage{xcolor}
\usepackage[hidelinks]{hyperref}

\newcommand{\be}{\begin{equation}}
\newcommand{\ee}{\end{equation}}

\begin{document}

\title{Observation of a dipolar quantum gas with metastable supersolid properties}

\author{L. Tanzi}
\affiliation{CNR-INO, S.S. A.~Gozzini di Pisa, via Moruzzi 1, 56124 Pisa, Italy}
\affiliation{LENS and Dip. di Fisica e Astronomia, Universit$\grave{\rm a}$ di Firenze, 50019 Sesto Fiorentino, Italy}
\author{E. Lucioni}
\affiliation{CNR-INO, S.S. A.~Gozzini di Pisa, via Moruzzi 1, 56124 Pisa, Italy}
\affiliation{LENS and Dip. di Fisica e Astronomia, Universit$\grave{\rm a}$ di Firenze, 50019 Sesto Fiorentino, Italy}
\author{F. Fam\`{a}}
\affiliation{CNR-INO, S.S. A.~Gozzini di Pisa, via Moruzzi 1, 56124 Pisa, Italy}
\author{J. Catani}
\affiliation{LENS and Dip. di Fisica e Astronomia, Universit$\grave{\rm a}$ di Firenze, 50019 Sesto Fiorentino, Italy}
\affiliation{CNR-INO, S.S. Sesto Fiorentino, 50019 Sesto Fiorentino, Italy}
\author{A. Fioretti}
\affiliation{CNR-INO, S.S. A.~Gozzini di Pisa, via Moruzzi 1, 56124 Pisa, Italy}
\author{C. Gabbanini}
\affiliation{CNR-INO, S.S. A.~Gozzini di Pisa, via Moruzzi 1, 56124 Pisa, Italy}
\author{R. N. Bisset}
\affiliation{Institut f\"ur Theoretische Physik, Leibniz Universit\"at Hannover, Appelstr. 2, 30167 Hannover, Germany}
\author{L. Santos}
\affiliation{Institut f\"ur Theoretische Physik, Leibniz Universit\"at Hannover, Appelstr. 2, 30167 Hannover, Germany}
\author{G. Modugno}
\affiliation{CNR-INO, S.S. A.~Gozzini di Pisa, via Moruzzi 1, 56124 Pisa, Italy}
\affiliation{LENS and Dip. di Fisica e Astronomia, Universit$\grave{\rm a}$ di Firenze, 50019 Sesto Fiorentino, Italy}
\date{\today}

\begin{abstract}
The competition of dipole-dipole and contact interactions leads to exciting new physics in dipolar gases, well-illustrated by the recent observation of quantum droplets and rotons in dipolar condensates. We show that the combination of the roton instability and quantum stabilization leads under proper conditions to a novel regime that presents supersolid properties, due to the coexistence of stripe modulation and phase coherence. In a combined experimental and theoretical analysis, we determine the parameter regime for the formation of coherent stripes, whose lifetime of a few tens of milliseconds is limited by the eventual destruction of the stripe pattern due to three-body losses. Our results open intriguing prospects for the development of long-lived dipolar supersolids.
\end{abstract}


\maketitle


Superfluidity and crystalline order are seemingly mutually exclusive properties. However, rather counterintuitively, both properties may coexist, resulting in an intriguing new phase known as a supersolid~\cite{Andreev1969,Legget1970,Chester1970,Boninsegni2012}. Although proposed $50$ years ago in $^4$He research, its experimental realization remains to this date elusive~\cite{Kim2012}. Recently, the idea of supersolidity has been revisited in the context of ultra cold atoms. The coexistence of phase coherence and density modulation has been reported in Bose-Einstein condensates~(BECs) in optical cavities~\cite{Leonard2017} and in the presence of synthetic spin-orbit coupling~\cite{Li2017}. The modulation in these systems is, however, infinitely stiff, since it is externally imposed.


This stiffness may be overcome in dipolar gases~\cite{Lahaye2009,Baranov2012}, allowing the realization of supersolids genuinely
resulting from interparticle interactions. Two recent experimental developments open exciting perspectives in this direction. On the one hand, the interplay of anisotropic dipole-dipole interactions, isotropic contact interactions and external confinement may lead to the appearance of a dispersion minimum that resembles the celebrated Helium roton~\cite{Santos2003}. Dipolar rotons have been observed very recently in experiments with Erbium atoms~\cite{Chomaz2018,Petter2018}. Interestingly, by decreasing the $s$-wave scattering length, the roton gap may be easily reduced until it vanishes, resulting in the so-called roton instability~\cite{Santos2003}. Although in the absence of stabilizing forces such an instability results in local collapses~\cite{Komineas2007}, a repulsive force at short range could stabilize a supersolid~\cite{Macia2012,Lu2015}. Interestingly, such a stabilization mechanism may be provided by quantum fluctuations~\cite{Petrov2015}, whose role is dramatically enhanced by the competition of dipole-dipole and contact interactions~\cite{Wachtler2016,Saito2016,Baillie2016,Bisset2016}. Quantum stabilization results in the formation of stable quantum droplets, as recently observed in a series of remarkable experiments~\cite{Kadau2016,Ferrier2016,Chomaz2016,Schmitt2016}, also demonstrating their self-bound nature~\cite{Schmitt2016}. In the presence of a trap, regular arrays of multiple droplets form due to dipolar repulsion~\cite{Kadau2016,Ferrier2016,Wenzel2017}. They however lack the necessary coherence to establish a supersolid phase, due to the weak tunneling between neighbouring droplets~\cite{Wenzel2017,Baillie2017}. Very recently, it has been predicted that stationary states of a dipolar Bose gas may acquire supersolid characteristics under the appropriate conditions \cite{Baillie2017,Roccuzzo2018}.


In this Letter, we show that ramping through the rotonic instability~\cite{Chomaz2018} in a weakly confined, strongly dipolar Dysprosium BEC results in the formation of a metastable, coherent stripe modulation, in a narrow range of scattering lengths close to the instability. By means of a combined experimental and theoretical investigation, we study the dynamics of the emerging density modulation, identifying the novel coherent regime as an array of weakly-bound droplets with a significant overlap and well-defined phase relation. Hence the stripes present supersolid properties, although they have a finite lifetime due to three-body losses. The stripe regime has distinct properties from the incoherent regime appearing for lower scattering length, which we identify as arrays of strongly-bound, but rapidly-decaying droplets~\cite{Ferrier2016,Wenzel2017}. Our results open exciting perspectives for the realization of long-lived dipolar supersolids.



Our experiment is based on a BEC of $^{162}$Dy atoms, with typical atom number $N$=4$\times$10$^4$ and undetectable thermal component \cite{Lucioni2017,Lucioni2018}. Two crossed optical potentials create a trap with frequencies $\omega_{x,y,z}$=2$\pi$(18.5, 53, 81)~Hz. A homogeneous magnetic field $B$ aligns the dipoles along the $z$ axis. Dy atoms in their ground state have a dipolar length $a_{dd}$=$\mu_0\mu^2 m/12\pi\hbar^2$$\simeq$130~$a_0$, for mass $m$ and dipole moment $\mu$. The s-wave scattering length $a_s$ is controlled via a magnetic Feshbach resonance located around 5.1~G \cite{Tang2016,Suppl}. The condensate is initially created with $a_s$ close to the background value $a_{bg}$=157(4)~$a_0$ \cite{Tang2016}. The magnetic field is then changed slowly in time to decrease $a_s$, with a final ramp from a stable BEC at $a_s$=108 $a_0$ into the roton instability \cite{Suppl}.



\begin{figure}[t]
\includegraphics[width=\columnwidth]{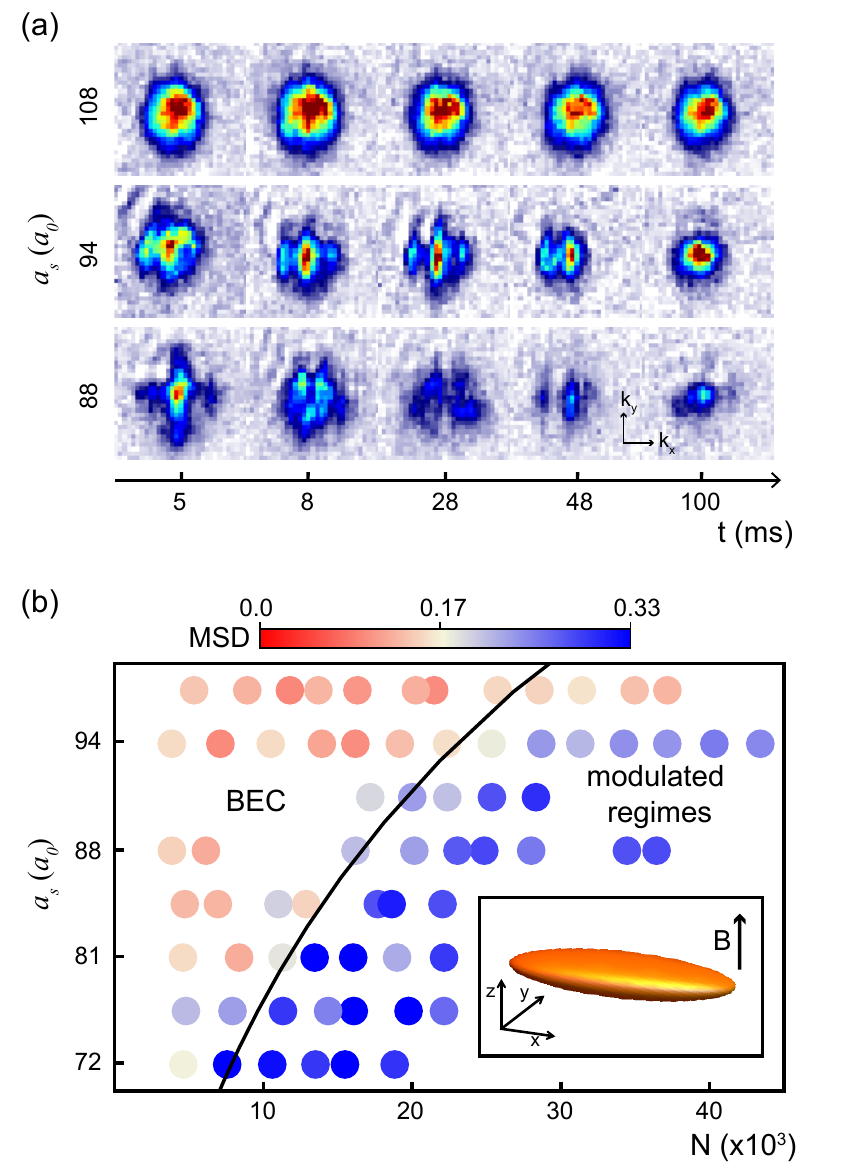}
\caption{
a) Typical momentum distribution $n(k_x,k_y)$ for different evolution times in three regimes: (top) $B$=5.305~G~($a\simeq108~a_0$, BEC regime); (middle) $B$=5.279~G~($a\simeq94~a_0$, stripe regime); (bottom) $B$=5.272~G~($a\simeq88~a_0$, incoherent regime).
b) Mean squared deviation (MSD) of $n(k_x,k_y)$ from a Gaussian, distinguishing BEC and stripe/incoherent regions in the $B$-$N$ plane~\cite{Suppl}. The black line represents the theoretical prediction for the roton instability~\cite{Chomaz2018,Suppl}.
Inset: trap geometry.}
\label{fig1}
\end{figure}



Our observable is the density distribution after 62 ms of free expansion at the final $a_s$, detected by absorption imaging along the $z$ direction. We interpret it as the momentum distribution $n(k_x,k_y)$ in the ($x$,$y$) plane~\cite{Suppl}. As a function of the final $a_s$, we observe three distinct regimes. For large $a_s$ the condensate does not qualitatively change compared to the initial BEC~(BEC regime). At intermediate $a_s$ the BEC develops a stripe-like modulation, but global phase coherence is preserved~(stripe regime). Finally for low $a_s$ global phase coherence and stripe regularity are rapidly lost~(incoherent regime). In the following we characterize in detail these regimes.

Figure~\ref{fig1}(a) shows time-of-flight pictures for three different final $a_s$ at different hold times $t$ after the end of the ramp. The upper panel, for $a_s\simeq108~a_0$, illustrates the BEC regime. As the scattering length is lowered to $a_s\simeq94~a_0$, for a limited range of scattering lengths, a stripe modulation spontaneously emerges~(middle panel): the momentum distribution shows small side peaks along the weak trap axis, with characteristic momentum $\bar{k}_x$=1.2(2)~$\mu$m$^{-1}$, close to the roton momentum predicted for an unconfined system at the instability, $k_{rot}$=1.53~$\mu$m$^{-1}$ \cite{Chomaz2018,Suppl}. The shape of $n(k_x,k_y)$ is reproducible from shot to shot, and is maintained for several tens of milliseconds. For longer times ($t\gtrsim100$~ms) an unmodulated BEC is recovered. For smaller $a_s$ values (bottom panel), $n(k_x,k_y)$ presents structures also along $k_y$, with maxima and minima distributed irregularly in the $(k_x,k_y)$ plane, as well as very large shot-to-shot variations. At longer times, we observe small condensates with large thermal fractions.

There is a marked dependence on $N$ of the critical scattering length at which we observe the onset of the modulated regimes. Figure~\ref{fig1}(b) shows the evolution of a phenomenological observable that quantifies the deviation of the momentum distribution from that of a BEC. One notes a region of small deviations for large $a_s$~(BEC regime), clearly separated from a region of large deviations for smaller $a_s$. The trend of the critical $a_s$ is reasonably well reproduced by numerical calculations based on the theory of roton instability~\cite{Chomaz2018,Suppl}.


\begin{figure}[t]
\includegraphics[width=\columnwidth]{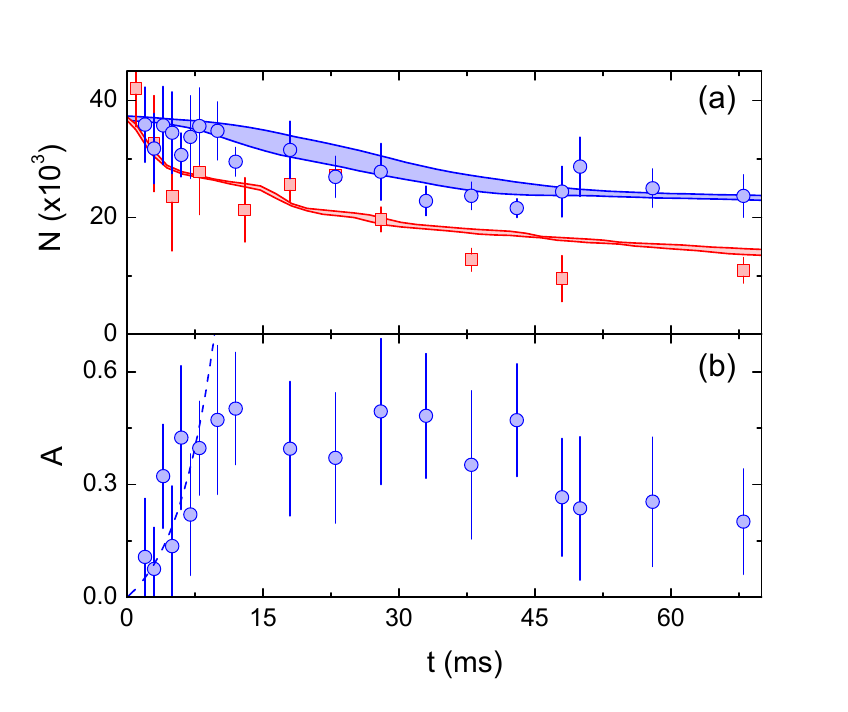}
\caption{
a) Time evolution of the atom number for stripes ($B$=5.279~G, blue circles) and incoherent ($B$=5.272~G, red squares) regimes. Blue and red shaded areas represent the atom loss predicted by our dynamical simulations at $a_s\simeq94~a_0$ and $a_s\simeq88~a_0$, respectively. b) Time evolution of the interference amplitude $A$ in the stripe regime ($B$=5.279~G). The dashed line is an exponential fit to the initial ($t\le$10 ms) growth. The error bars represent the standard deviation of about 40 measurements.
}
\label{fig2}
\end{figure}



The time evolution of the atom number $N(t)$ is shown in Fig.~\ref{fig2}a. Both stripe and incoherent regimes feature an initial loss on timescales much faster than the typical lifetime of a BEC at $B$=5.305~G, $\tau_{BEC}$$\simeq$500~ms \cite{Suppl}. We can estimate the in-situ mean density from the loss rate, since $\dot{N}/N=-L_3\langle n^2\rangle$, with $\langle n^2\rangle$ the mean quadratic density. Using the recombination constant measured from the decay of the stable BEC at $B$=5.305~G, $L_3=2.5(3)\times 10^{-28}$~cm$^{6}$s$^{-1}$, we estimate a similar mean density of order $n\simeq 5\times 10^{14}$~cm$^{-3}$, for both stripe and incoherent regimes~\cite{Suppl}. This is about $10$ times larger than the calculated BEC density, suggesting that in both modulated regimes the Lee-Huang-Yang~(LHY) repulsion has a stabilizing role~\cite{Petrov2015,Wachtler2016,Saito2016,Baillie2016,Bisset2016,Ferrier2016}.



We analyze the stripe regime by fitting the $y$-averaged distributions $n(k_x)$ with a two-slit model $n(k_x)=C_0\exp(-k_x^2/2\sigma_x^2)[1+C_1\cos^2(\pi k_x/\bar{k}_x+\phi)]$. The interference amplitude $A$ is defined as the relative weight of the side peaks in $n(k_x)$ with respect to the central one \cite{Suppl}, and provides information on the depth of the density modulation. The interference phase $\phi$ provides instead a measure of the robustness of the stripe pattern, both in what concerns the phase locking between the stripes and their relative distances.


\begin{figure}[t]
\includegraphics[width=\columnwidth]{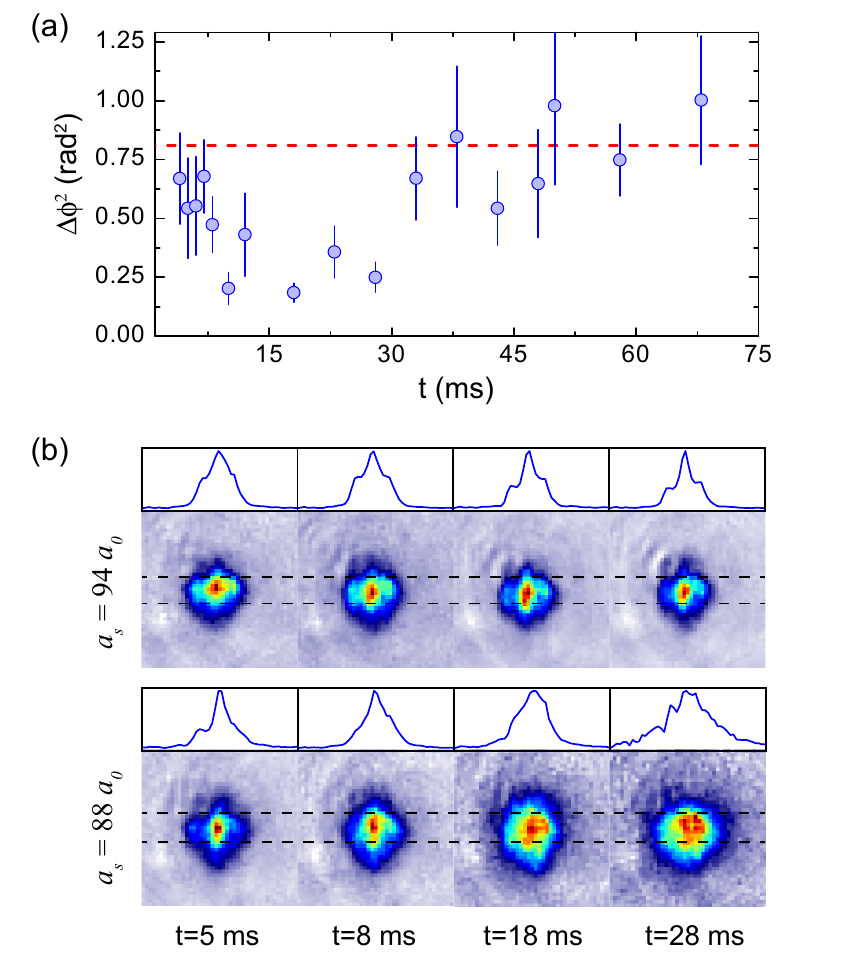}
\caption{
a) Time evolution of the interference phase variance $\Delta\phi^2$ in the stripe regime ($a\simeq94~a_0$). The error bars correspond to $\Delta\phi^2 2/(2N-2)$, with $N\simeq40$ the number of measurements for each dataset. The red-dashed line is the expected variance for a uniformly distributed phase. b) Averaged momentum distribution $\bar{n}(k_x,k_y)$ over 40 absorption images in the stripe regime (top) and in the incoherent regime (bottom) at different evolution times. The profiles are obtained by integrating $\bar{n}(k_x,k_y)$ along $k_y$ in the region between dashed lines.
}
 \label{fig3}
\end{figure}


Figure~\ref{fig2}b shows the evolution of $A$ in the stripe regime. The initial exponential growth is consistent with the onset of the roton instability observed in previous experiments~\cite{Chomaz2018}. After this initial growth, $A$ remains approximately constant for about $30$~ms and then decreases. The reduction of $A$ at longer times gives evidence of the progressive disappearance of the stripe modulation, compatible with the reduction of the atom loss rate observed in Fig.~\ref{fig2}a.


\begin{figure}[t]
\includegraphics[width=\columnwidth]{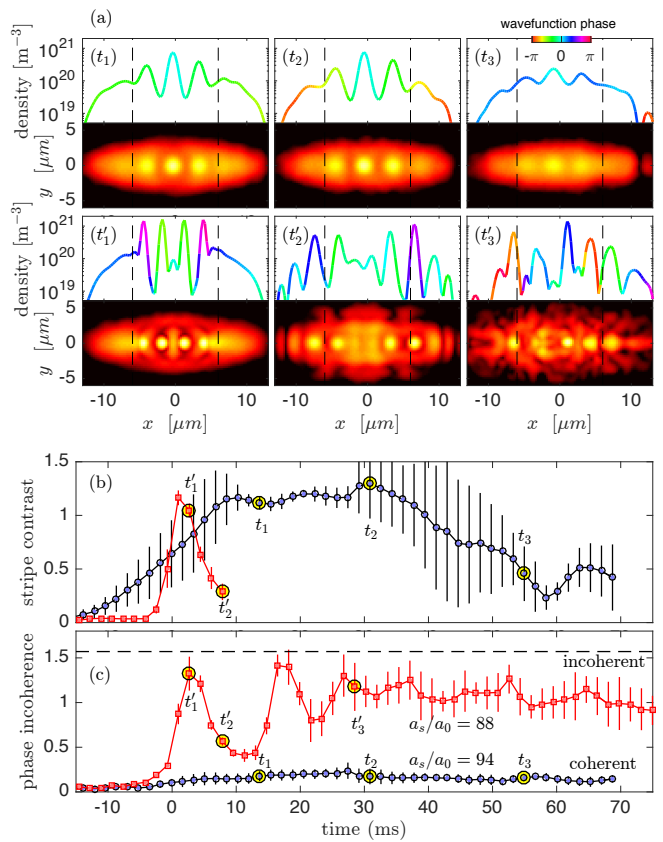}
\caption{
Simulations of stripe formation and evolution. (a) Snapshots for a simulation with coherent stripes after a ramp to $a_s/a_0=94$ are shown at times $(t_1,t_2,t_3)=(13.7,30.9,55)$ ms, while the incoherent regime can be seen in another simulation after a ramp to $a_s/a_0=88$ for $(t_1^\prime,t_2^\prime,t_3^\prime)=(2.7,7.9,28.5)$ ms. Density cuts $n(x,0,0)$ -- with color representing wavefunction phase -- and column densities $\int dz n(x,y,z)$ are shown.  (b) Stripe contrast $\mathcal{C}$ for stable stripes with $a_s/a_0=94$ (blue circles) and unstable stripes with $a_s/a_0=88$ (red squares). The error bars represent standard deviations for six simulations, each with different initial noise. (c) Phase incoherence, $\alpha^I=\int^\tau dxdy n|\alpha - \langle \alpha \rangle |/\int^\tau dxdy n $, where $\alpha(x,y,0)$ is the wavefunction phase, $\langle\alpha\rangle$ is its average over $\tau$, defined as the region between the dashed lines in (a) \cite{Suppl}. The limit $\alpha^I=0$ indicates global phase coherence while $\alpha^I=\pi/2$ signals incoherence.
}
\label{fig4}
\end{figure}



Figure~\ref{fig3} shows the key observations for the coherence. Fig.~\ref{fig3}a depicts the time evolution of the variance $\Delta\phi^2$ in the stripe regime, obtained from about $40$ realizations for each evolution time. At the initial stages of the rotonic instability, we observe a large variation of $\phi$, which may be explained due to shot-to-shot differences in the quantum and thermal seeding of the instability that lead to a marked variation, for a fixed time, of $A$. Remarkably, we observe that after the stripe formation~(first $10$~ms), $\Delta\phi^2$ remains small for approximately $20$~ms, revealing that the stripes remain stable and coherent for a time significantly longer than their formation time. After this time, $\Delta\phi^2$ increases, eventually reaching the expectation value for a uniformly distributed $\phi$, corresponding to a fully incoherent or disorganized stripe pattern.

When decreasing the final $a_s$, the system enters the incoherent regime. In order to compare stripe and incoherent regimes, we study their average momentum distribution over approximately $40$ absorption images at different evolution times, see Fig.~\ref{fig3}b. The persisting side peaks in the stripe regime confirm the existence of a stable coherent stripe pattern. In contrast, when $a_s$ is reduced into the incoherent regime, side peaks are only visible during the pattern formation~($t=5$~ms), whereas already at $t=18$~ms no clear pattern is recognizable, showing that coherence and/or pattern stability is quickly lost after the instability develops.



In direct support of our experiments, we have performed realistic 3D simulations of the dynamics during and after the ramp of $a_s$ using the generalized Gross-Pitaevskii equation, which includes the stabilizing effects of quantum fluctuations \cite{Wachtler2016,Bisset2016,Ferrier2016,Chomaz2016}.
We also seed the initial states with quantum and thermal fluctuations according to the Truncated-Wigner prescription, include three-body losses, and the $a_s(B)$ dependence that, within the experimental uncertainty, provides the best experiment-theory agreement~\cite{Suppl}.
The simulations support the experimental observations and provide key insights into the nature of both stripe and incoherent regimes. First of all, they confirm that the observed stripe regime is triggered by a roton instability, similar to the one observed in an Er system \cite{Chomaz2018}, but in our experiments evolves into a long-lived density modulation due to the stabilizing role of quantum fluctuations.
This is shown for example by the good agreement of theory and experiment for $N(t)$ in both the stripe and incoherent regime (see Fig.~\ref{fig2}a), which confirms that the atom number decay is due to the appearance of high-density modulations, where the LHY energy plays a significant role.

The most important results of the simulations are summarized in Figs.~\ref{fig4}. They confirm a marked difference between the observed stripe and incoherent regimes. Since simulating the dynamics during the free expansion is challenging \cite{Chomaz2018}, we study the in-trap density and phase distributions.
The wavefunction density/phase plots in Figs.~\ref{fig4}a show that the density modulation in the two regimes has a different nature. In the stripe regime (Fig.~\ref{fig4}a, top panels), the modulation originates from the formation of an array of weakly-bound droplets along the $x$ direction, on top of a sizeable BEC background; the modulation slowly decays in time due to three-body losses and eventually a moderately excited BEC is recovered at long times. The phase $\alpha$ of the wavefunction remains approximately uniform during the whole time evolution, apart from a small parabolic phase profile that corresponds to an axial breathing mode excited by the changing density distribution at the initial instability.
In the incoherent regime, in the absence of three-body losses, our numerics predicts the formation of an array of tightly-bound droplets \cite{Suppl}. In contrast to the stripe regime, they would present no significant overlap (see e.g. Fig.~\ref{fig4}a, bottom left), and hence would rapidly become incoherent \cite{Wenzel2017}. However, in the presence of the experimental losses (Fig.~\ref{fig4}a, bottom panels), although tightly-bound droplets develop initially, the larger peak density causes their very rapid decay before they can reach an equilibrium situation. The droplet decay results in strong excitations that cause violent density fluctuations in both the $x$ and $y$ directions. These density fluctuations result in the irregular, incoherent patterns that we observe experimentally after the free expansion.

We have numerically studied the growth of the density modulation as well as the phase profile, averaging over different realizations~(characterized by different initial fluctuations). For the stripe regime, the calculated stripe contrast $\mathcal{C}$ in~Fig.~\ref{fig4}b -- defined as the amplitude of the stripe density oscillations divided by the amplitude of an overall Thomas-Fermi fit \cite{Suppl} -- is in good agreement with the experimental observable $A$~(Fig.~\ref{fig2}b). An initial growth during approximately $10$~ms is followed by a plateau for $30$~ms, and a later decay towards zero. The apparent longer growth time in the simulations arises because the momentum space observable $A$ at the beginning of the pattern growth depends quadratically on the position space quantity $\mathcal{C}$.
Figure~\ref{fig4}c shows the phase incoherence $\alpha^I$ (defined in the caption of ~Fig.~\ref{fig4}), after removing the parabolic phase profile due to the breathing oscillation~\cite{Suppl}. Remarkably, the spatial variation of the phase remains very small during the whole evolution, indicating the presence of a robust phase locking of the stripes. The numerically observed formation of coherent stripes is in agreement with the small $\Delta\phi^2$ of Fig.~\ref{fig3}a. In contrast, the phase variation is very large in the incoherent regime~(the observed modulation in $\alpha^I$ is given by the nucleation and unraveling of unstable droplets). Note that, since the interference phase is sensitive to both wavefuction phase and stripe stability, we accordingly observe strong fluctuations of $\phi$ (Fig.~\ref{fig3}a) when $\mathcal{C}$  fluctuates (Fig.~\ref{fig4}b), although $\alpha$ is still coherent. We attribute the contrast fluctuations around 30-50 ms, when $A$ is still large, to an effect of three-body losses.

The novel stripe regime hence reveals supersolid properties, due to the co-existence of phase coherence and density modulation. In the absence of losses, our numerics reveals the formation of stable coherent stripes, which would still be in an excited state as a result of crossing the first-order phase transition when ramping down the scattering length \cite{Suppl}. However, three-body losses render the stripe pattern eventually unstable in our experiments, with a life-time of approximately $30$~ms. This instability is however not related with the loss of phase coherence, since the latter remains high at any time despite quantum and thermal phase fluctuations, three-body losses, and the breathing oscillation~(Fig.~\ref{fig4}c). Our analysis shows that the finite life-time of coherent stripes rather results from the eventual instability of the stripe modulation~(Fig.~\ref{fig4}b), which leads to the experimentally observed time dependence of both $A$ and $\Delta\phi^2$. Once the density modulation vanishes the system remains highly coherent, in agreement with our experimental observation at long times of a large BEC, in stark contrast to our observation in the incoherent regime~(Fig.~\ref{fig1}a)~\cite{Suppl}.



In summary, we report a novel regime in a dipolar quantum gas, formed by overlapping weakly-bound droplets, that exists in a narrow range of scattering lengths close to the roton instability. Due to its simultaneous phase coherence and density modulation, this regime exhibits the properties of an excited, metastable supersolid. Whereas the excitation occurs in any case due to the crossing of a first-order phase transition, the observed metastability stems from three-body losses, which in our case limit the lifetime to approximately $30$~ms. Longer lifetimes might be achieved by searching magnetic-field regions in Dy isotopes with lower loss rates, or going to larger scattering lengths using larger atom number and less confining traps. Longer lifetimes will be important to test the stripe superfluidity, which is a requisite to assess their supersolid nature~\cite{Andreev1969,Legget1970,Chester1970,Boninsegni2012}.

{\it Note added:} We recently became aware of a complementary experiment~\cite{Boettcher2019}, which was motivated by our initial experimental observations. The new theoretical analysis of this revised version confirms and complements their numerical results.



We acknowledge funding by the EC-H2020 research and innovation program (Grant 641122 - QUIC) and the DFG (SFB 1227 DQ-mat and FOR2247).
RNB was supported by the European Union's Horizon 2020 research and innovation programme under the Marie Sk\l odowska-Curie grant agreement No. 793504 (DDQF).
We acknowledge support by L. A. Gizzi, S. Gozzini and M. Inguscio, useful conversations with M. Fattori, G. Ferrari and S. Stringari, and technical assistance by A. Barbini, F. Pardini, M. Tagliaferri and M. Voliani.



\newpage

\section*{Supplementary material}

\subsection{Magnetic field dependence of the scattering length}
\begin{figure}
\includegraphics[width=\columnwidth]{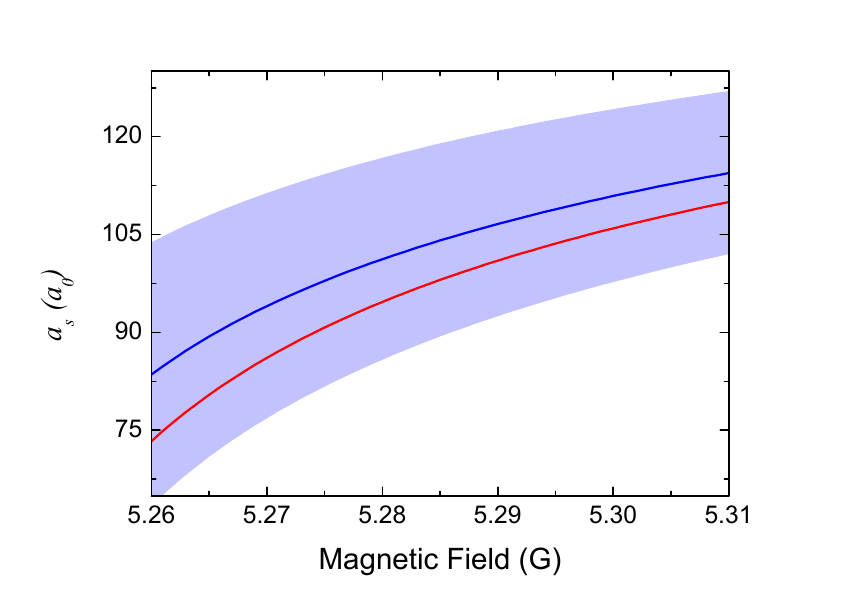}
\caption{Contact scattering length versus magnetic field. The blue-solid line represents our best estimate for $a_s(B)$ based on the Feshbach resonances parameters. The blue region sets the limits of confidence for $a_s(B)$, given the experimental uncertainty on the resonances parameters. The red-solid line is the conversion of $a_s(B)$ used throughout the paper, obtained by comparing experimental and theoretical observations.} \label{fig_asc}
\end{figure}

In the experiment the magnetic field is calibrated through radio-frequency (rf) spectroscopy between two hyperfine states at B=5.305~G, in the stable region. The line-width of the rf transition is 3~kHz, corresponding to a systematic uncertainty on the magnetic field of about 1~mG.

We tune the contact scattering length $a_s(B)$ using a set of three Feshbach resonances located at $B_1=5.145(3)$~G, $B_2=5.231(3)$~G, and $B_3=5.244(3)$~G, respectively. We estimate positions and widths for each Feshbach resonance using both loss spectroscopy and thermalization measurements. The widths of the first and second Feshbach resonances are $\Delta B_1=32(7)$~mG, $\Delta B_2=8(3)$~mG, respectively. The third resonance has a width of the order of 1~mG, and does not affect the contact scattering properties in the magnetic field range of interest. The blue region in Fig.\ref{fig_asc} shows the limits of confidence for our best $a_s(B)$ estimate (blue-solid line), assuming a background scattering length $a_{bg}=157(4)~a_0$ \cite{Tang2016}. 

The dynamical evolution of the system is studied experimentally at three specific magnetic fields: B=5.305~G (BEC regime), B=5.279~G (stripe regime), B=5.272~G (incoherent regime). 
Due to the large experimental uncertainty on $a_s(B)$, we identify a precise $B$ to $a_s$ conversion by comparing experimental and numerical data. In particular, we compare the instability onset at fixed atom number N=$4\times10^4$ (occurring at B=5.279~G in the experiment and at $a_s=94~a_0$ in the numerics). The conversion that provides the best experiment-theory agreement is: $a_s(B)=\big(1-\frac{0.032}{B-5.145}\big)\big(1-\frac{0.01}{B-5.231}\big)157a_0$, well inside our region of confidence, see red line in Fig.\ref{fig_asc}.

\subsection{Three-body recombination}

\begin{figure}
\includegraphics[width=\columnwidth]{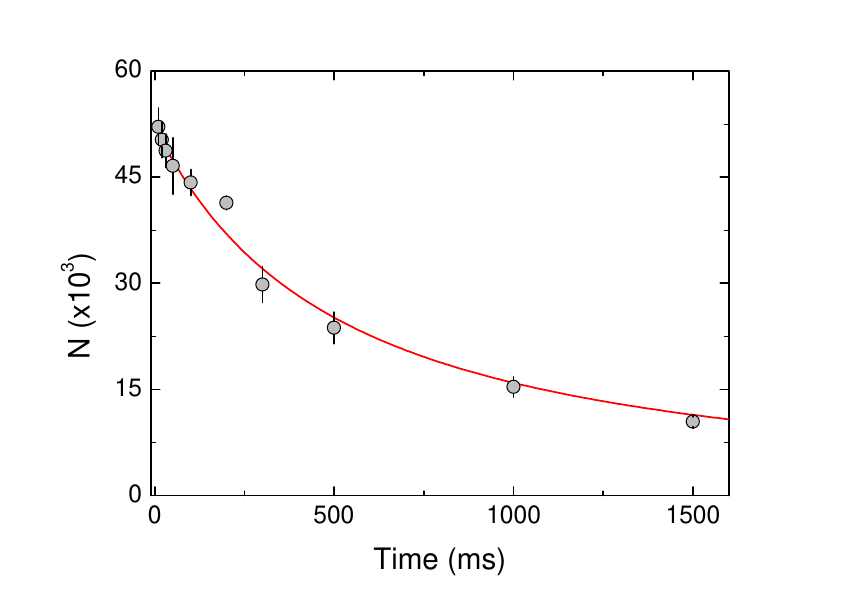}
\caption{Time evolution for the atom number for a BEC confined in an optical dipole trap with frequencies $\omega=2\pi(18.5,51,83)$ Hz, at $B=5.305$~G. The red-solid line is a fit to the data points, using the numerical solution valid for our specific trap geometry, see text.}  \label{fig_BEC}
\end{figure}

We attribute the decay of the atom number observed experimentally in the modulated regimes to three-body recombination. Since the three-body recombination rate $L_3$ is not known for $^{162}$Dy, we have determined its value experimentally, using two distinct methodologies. In a first set of measurements, we have studied the time evolution of the atom number of a BEC confined in standard dipole trap with frequencies $\omega=2\pi(18.5,51,83)\pm(2)$~Hz, at $B$=5.305~G, in the stable regime. This is related to $L_3$ via the equation: $\dot{N}/N=-L_3 \langle n^2\rangle$. Numerical integration of this equation gives an analytical form for $N(t)$, which can be used to fit the experimental points, leaving $L_3$ as the only free-parameter. By evaluating the mean density within the Thomas-Fermi approximation in our specific trap geometry, we obtain $N(t)=5^{5/4}\Big(\frac{5 + 4 N_0^{4/5} (1.15\times10^{18})^2 L_3  t}{N_0^{4/5}}\Big)^{-5/4}$, where $N_0$ is the initial atom number. In Fig.\ref{fig_BEC} we show the result of the fit of $N(t)$ to our data. From this analysis, we obtain $L_3=2.5(3)\times10^{-28}$~cm$^6$/s. This value is larger than the values measured for $^{164}$Dy, which are of the order of $10^{-29}$ ~cm$^6$/s \cite{Ferrier2016}.

In a second set of experiments, we have used a thermal cloud confined in a deep optical dipole trap with mean frequency $\bar{\omega}/2\pi=88(5)$~Hz. We have then recorded the time evolution of its atom number and temperature. By fitting our measurements with a set of coupled equations for $\dot{N}$ and $\dot{T}$ (see ref. \cite{Roy2013}), we obtain $L_3^{th}=6L_3=1.3(2)\times10^{-27}$~cm$^6$/s, in agreement with the BEC measurements.

\begin{figure}
\includegraphics[width=\columnwidth]{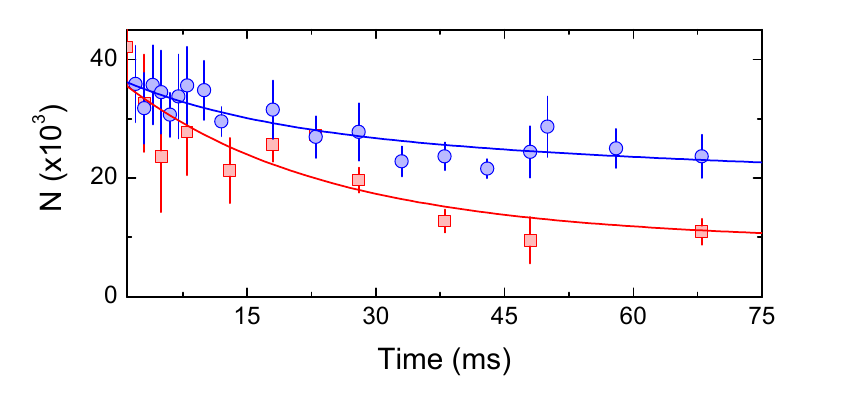}
\caption{Time evolution of the atom number for stripe ($B$=5.279~G, blue circles) and incoherent regimes ($B$=5.272~G, red squares). Lines are exponential fits to data, see text.}  \label{fig_N}
\end{figure}

Fig.\ref{fig_N} shows the atom number decay for stripe and incoherent regimes (blue circles and red squares, respectively). By fitting both datasets with the simplified model $N(t)=\Delta N\exp(-t/\tau)+N_0$ on the entire time span, we extract similar decay rates for stripe and incoherent regimes, $\tau$=23(12)~ms and $\tau$=24(10)~ms respectively. We use the recombination constant determined above to estimate the mean density for both regimes, $n\simeq 5\times 10^{14}$~cm$^{-3}$, about a factor 10 larger than the calculated mean density of the BEC. In the case of the incoherent regime, the numerical simulation for $N(t)$ reported in Fig.1 show faster losses during the first 10 ms, associated to the formation of denser droplets than in the stripe regime. Such rapid loss in consistent with the experimental data, but cannot be reproduced by the simplified exponential loss model employed here.

\subsection{Magnetic field ramps}
The atoms are condensed in a crossed optical dipole trap made of a Gaussian beam with waist 41~$\mu$m, and an elliptical beam with horizontal (vertical) waist 81(36)~$\mu$m. The final trap frequencies are the ones used in the experiments, $\omega=2\pi(18.5,51,83)\pm2$~Hz. The condensate is initially created at $B$=5.5~G, where $a_s$ is close to the background value $a_{bg}$=157(4)~$a_0$. The magnetic field is subsequently changed to $B$=5.305~G ($a_s\simeq$108~$a_0$, according to our calibration of $a_s$) with a linear ramp in 80~ms. We use a second 30~ms linear ramp to enter the unstable regimes, reaching a final value of the magnetic field ranging in the interval $B$=(5.26-5.29)~G. After variable waiting time at the final magnetic field, we switch off all the dipole traps and we take an absorption image of the atoms after 62~ms of free fall. 

\subsection{Fitting procedures for the experimental momentum distribution}
\begin{figure}
\includegraphics[width=\columnwidth]{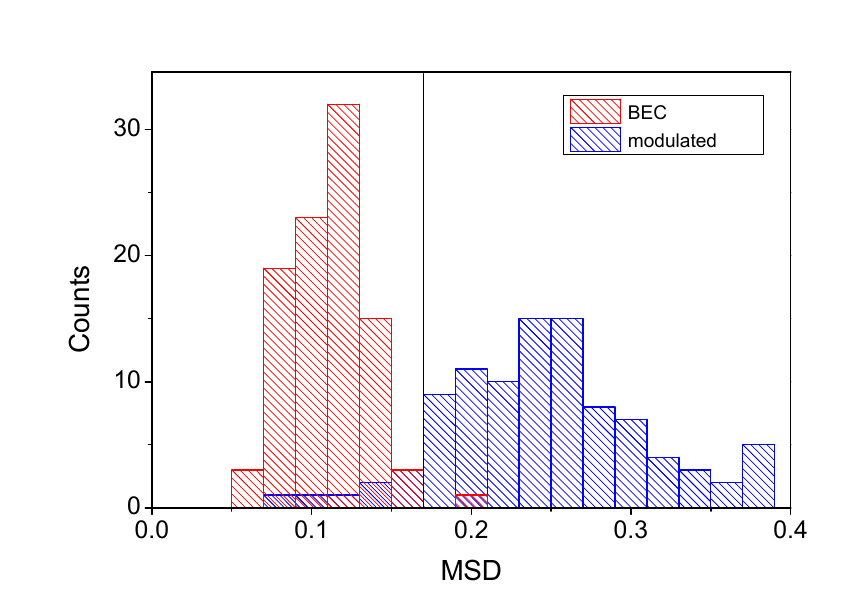}
\caption{Histogram of the mean squared deviation MSD for BEC (red) and modulated regimes (blue).}  \label{fig_fig1h}
\end{figure}

\begin{figure}
\includegraphics[width=\columnwidth]{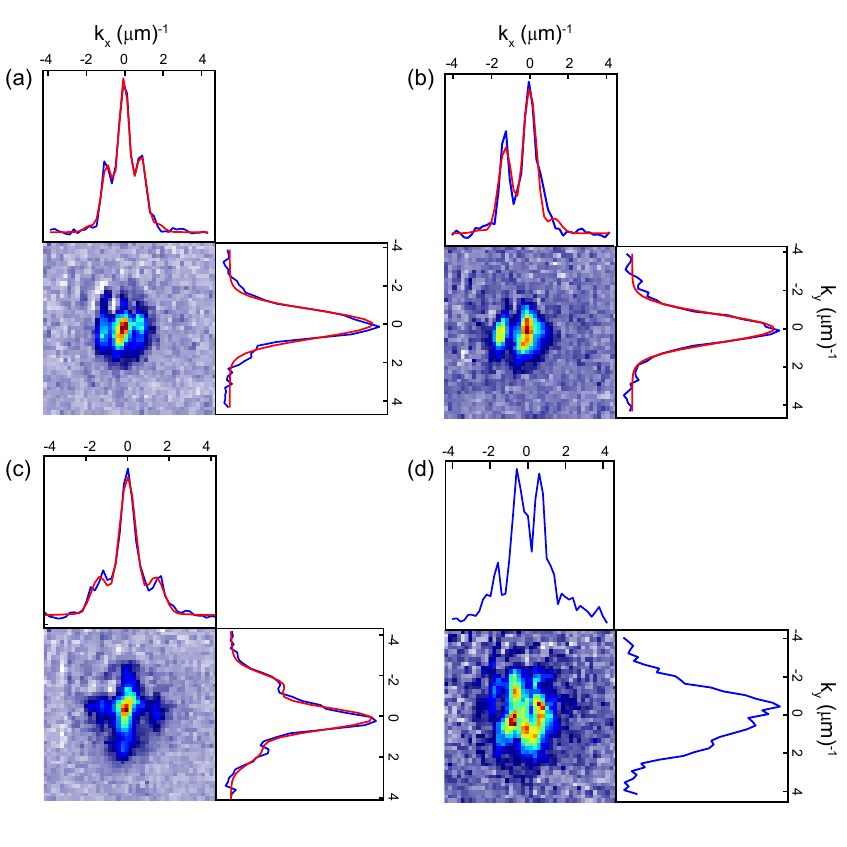}
\caption{Examples of raw momentum distributions and corresponding integrated profiles along $x$ and $y$ (blue lines). The red lines are fits to the integrated profiles with Eq. \ref{eq:doubleslit}. The distributions correspond to the following parameters: a) $a_s\simeq94~a_0$, $t$=18~ms; b) $a_s\simeq94~a_0$, $t$=43~ms; c) $a_s\simeq88~a_0$, $t$=5~ms; d) $a_s\simeq88~a_0$, $t$=18~ms.}  \label{fig_prof}
\end{figure}

In this section we describe the strategies adopted for analyzing the individual 2D experimental distributions. For each magnetic field and evolution time we recorded between 40 and 70 time-of-flight distributions, obtained by absorption imaging at the final scattering length. Our resolution in momentum space is 0.2~$\mu$m$^{-1}$ (1/e Gaussian width). We have verified that the time-of-flight distribution does not change qualitatively by quenching back the magnetic field to $B$=5.305~G (BEC regime) before the expansion.
Since we work in a regime dominated by dipolar interactions we cannot exclude interactions effects during the time-of-flight, thus affecting the momentum distribution. The characteristic momentum observed in the experiment, $\bar{k}_x$=1.2(2)~$\mu m^{-1}$, is smaller than both the theoretically calculated $k_{rot}$=1.53$~\mu m^{-1}$ (Sec.~\ref{Sec:Ana}), and the numerically calculated positions of the momentum side peaks immediately following the roton instability, $\bar{k}_x$=1.6~$\mu m^{-1}$ (Sec.~\ref{Sec:ISP}), the latter of which is related to the initial spatial frequency of the stripes. A possible explanation is a modification of the momentum distribution during the first phases of the free expansion. Although such effect seemed not to be present in Er \cite{Chomaz2018}, we cannot exclude it for the more dipolar Dy systems. The assessment of this phenomenon will be the subject of a future work.

The atom number is determined independently of any fitting procedure, by evaluating the zeroth moment of the distributions $N=\sum_{x,y}n_c(x,y)$, where $n_c(x,y)$ is the $(x,y)$ pixel column density. From each distribution we extract two 1D profiles, $n(k_{x})$ and $n(k_{y})$ by integrating over the vertical and horizontal direction, respectively.

Fig.1(b) in the main text is constructed performing a Gaussian fit on $n(k_{x})$ at fixed evolution time of the order of the stripe formation time, $t$=8~ms. We evaluate the normalized mean squared deviation between raw data $n(k_{x})$ and the fitted Gaussian $g(k_{x})$ as: MSD=$\int [n(k_{x})-g(k_{x})]^2dk_x/\int g(k_{x})dk_x$. Fig.\ref{fig_fig1h} displays the MSD histogram both for a dataset with only BEC samples, realized at $B=5.305$~G, and a dataset with only modulated samples, realized at $B=5.26$~G and $B=5.263$~G. The discriminator between the two regimes is then recognized when MSD$\sim$0.17. The color-scale normalization in Fig.1(b) is defined in order to highlight such value of MSD.

The modulated regime is further analyzed using a double slit fit on $n(k_{x})$ and $n(k_{y})$:
\begin{equation}
n(k_i) = C_0 e^{-\frac{(k_i-k_{0})^2}{2\sigma^{2}}}\bigg[1+C_1\cos^2\Big((k_i-k_0)\frac{\pi}{\bar{k}_i}+\phi\Big)\bigg],
\label{eq:doubleslit}
\end{equation}
with $i=x,y$. In this analysis, all parameters are left unbounded. We only impose loose constrains on $\bar{k}_i$ for discarding unrealistic values. From this analysis we get the parameters discussed in the main text: the characteristic periodicity $\bar{k}_i$, the interference phase $\phi$, the envelop size $\sigma$, and the unmodulated and modulated amplitudes $C_0$ and $C_1$.
In the stripe regime we get information on the interference amplitude, reconstructing $n(k_x)$ for $\phi=0$ and fitting it with three Gaussians:
\begin{equation}
n(k_x)=A_0e^{-\frac{k_x^2}{2\sigma^2}}+A_1\bigg(e^{-\frac{(k_x-\bar{k}_x)^2}{2\sigma^2}}+e^{-\frac{(k_x+\bar{k}_x)^2}{2\sigma^2}}\bigg).
\label{eq:3Gauss}
\end{equation}
The interference amplitude $A$ is defined as the relative amplitude of side peaks with respect to the central peak, $A=A_1/A_0$. Examples of integrated profiles fitted with Eq. \ref{eq:doubleslit}, in both incoherent and stripe regimes are shown in Fig.\ref{fig_prof}.

\begin{figure}
\includegraphics[width=\columnwidth]{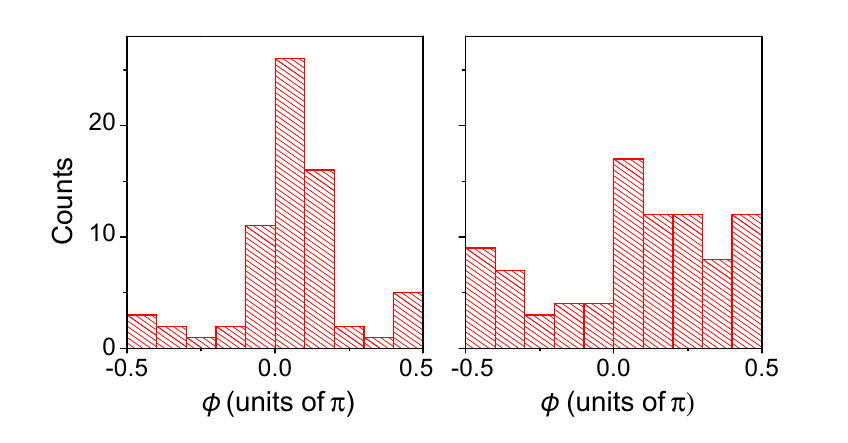}
\caption{Histograms of the interference phase $\phi$} in the stripe regime ($a=94~a_0$), for two evolution times, $t$=28~ms (left) and $t$=68~ms (right).  \label{fig_hist}
\end{figure}

As discussed in the main text, the interference phase $\phi$ provides a measure of the robustness of the stripe pattern, in what concerns both the stripe phase locking and their relative distances. For each evolution time in Fig.3 we study the distribution of $\phi$ for more than 40 images. Fig.\ref{fig_hist} shows two examples of the interference phase distribution, measured at different evolution times for $a\simeq94~a_0$ G. Up to approximately 30~ms, the phase distribution is peaked around $\phi$=0, and its variance is small. Instead, for increasing evolution times the phase diffuses, tending to a uniform distribution.

\subsection{Temperature measurements}
\begin{figure}
\includegraphics[width=\columnwidth]{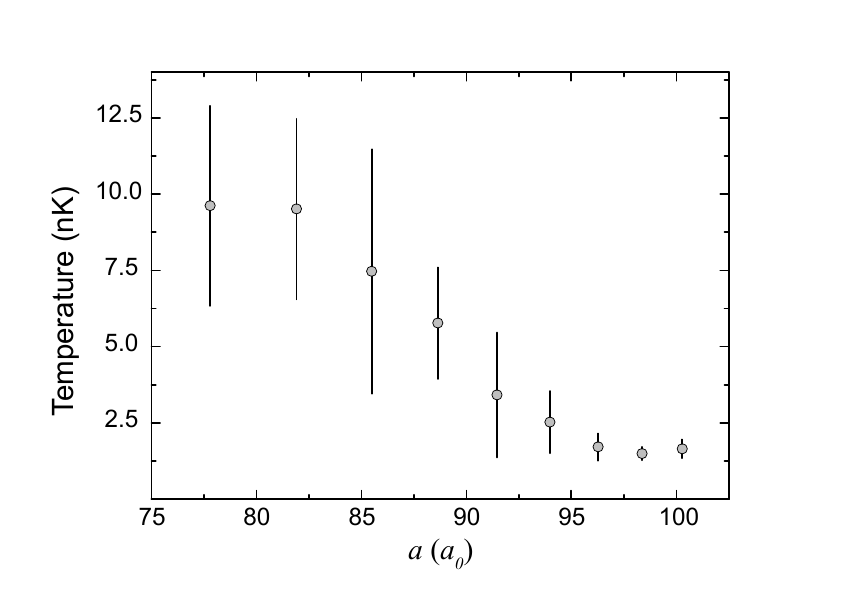}
\caption{Apparent temperature of the excited component after 200 ms evolution time at fixed magnetic field as a function of the scattering length.}  \label{fig_temp}
\end{figure}

Further evidence for the different degree of incoherence and excitation of the stripe and incoherent regimes can be obtained from an analysis of the apparent temperature of the system at relatively long times, after the density modulations have decayed and the system again appears as an ordinary BEC. Fig. \ref{fig_temp} reports the apparent temperature -- based on the width of the excited part of a bimodal fit -- at an evolution time $t$=200~ms, for different scattering lengths. The trap depth is $V_0$=60~nK. When the system is prepared in the BEC regime, we do not observe any visible thermal component in the momentum distribution. A fit with two unconstrained Gaussians is indeed unable to distinguish two components with different momentum widths (given the noise level, we estimate that we can detect a thermal component only if its relative population is larger than 30\%). In the stripe regime ($a_s\simeq94~a_0$), we instead detect two different components -- a condensate and an excited component, with approximately equal populations -- and we can associate a temperature of about 3~nK to the excited component. For smaller scattering lengths, after the system enters the incoherent regime ($a_s\le90~a_0$), a larger excited component appears, having a larger temperature up to 10~nK. We note that in both cases the excited component can be observed only at $t\approx$~200~ms, since at longer times it disappears, thermalizing with the BEC and perhaps also through evaporation from the trap.

\subsection{Theory of the roton instability}\label{Sec:Ana}
Ref. \cite{Chomaz2018} studied theoretically the dispersion relation of a dipolar BEC in the Thomas-Fermi limit, in a trap with negligible confinement along the long axis (in our case, the $x$ axis). Close to the roton instability, the dispersion has the form
\begin{equation}
\epsilon(k_x)^2=\Delta^2+\frac{2\hbar^2k_{rot}^2}{m}\frac{\hbar^2}{2m}(k_x-k_{rot})^2\,,
\end{equation}
where $\Delta$ is the roton gap and $k_{rot}$ is the roton momentum. The gap is defined as
\begin{equation}
\Delta=\sqrt{E_0^2-E_I^2}\,,
\end{equation}
where
\begin{equation}
E_0^2=\frac{4\pi\hbar^2a_{dd}}{m}n_0\frac{\hbar^2}{2m}\Big(\frac{1}{R_z^2}+\frac{1}{R_y^2}\Big)
\end{equation}
and
\begin{equation}
E_I=\frac{8\pi\hbar^2(a_{dd}-a_s)}{3m}n_0\,,
\end{equation}
with $R_y$ and $R_z$ the Thomas-Fermi radii of the condensate and $n_0$ its peak density. The roton momentum is
\begin{equation}
k_{rot}=\sqrt{2mE_I}\,,
\end{equation}
implying the relation
\begin{equation}
\frac{\hbar^2 k_{rot}^2}{2m}=\frac{8\pi\hbar^2 (a_{dd}-a_s) n_0}{3m}\,.
\label{instability}
\end{equation}
Eq.\ref{instability} can be interpreted as the balance of the cost in kinetic energy for creating the roton density modulation and the gain in interaction energy of the dipoles stacked along $z$ in the density peaks.

At the roton instability, $\Delta=0$ implies $E_I^2=E_0^2$. Considering the experimental case, in which a trap is present also along the $x$ direction, and given the relation between the peak density and the atom number, $n_0=15N/8\pi R_xR_yR_z$, the condition for the instability can also be written in the form
\begin{equation}
N=\frac{3}{10}\frac{a_{dd}}{(a_{dd}-a_s)^2}\frac{R_x(R_y^2+R_z^2)}{R_yR_z}\,.
\label{instabilityb}
\end{equation}
Eq. \ref{instabilityb} is clearly not self-consistent, given the original assumption of negligible confinement along $x$, so it cannot be used to derive exact predictions.

To find the critical $a_s$, we calculated the radii using a variational method \cite{Giovanazzi}, including the LHY energy term \cite{Lima2012}. At the instability, the roton momentum is $k_{rot}$=1.53~$\mu$m$^{-1}$. By extending the calculation to different values of $N$, we construct the line separating the BEC regime from the stripe regime in the $B$-$N$ plane shown in Fig.1 of the paper. In particular, we find that the factor $R_x(R_y^2+R_z^2)/R_yR_z$ in Eq.~(\ref{instabilityb}) is approximately independent of $a_s$ but scales approximately as $N^{1/5}$, therefore leading to a scaling close to $N\propto (a_{dd}-a_s)^{-5/2}$ (the actual exponent is -2.32(2)). 

\subsection{Dynamic simulations}

\subsubsection{Numerical details}

In direct support of our experiments we perform realistic 3D simulations of the dynamics during and after the ramp of $a_s$ using the generalized Gross-Pitaevskii equation \cite{Wachtler2016,Bisset2016,Ferrier2016,Chomaz2016},
\begin{equation}
i\hbar\frac{\partial\psi}{\partial t} = \Big[ H_0 + V_{\rm MF}(\mathbf{x}) + V_{\rm QF}(\mathbf{x}) -i\hbar\frac{L_3}{2}|\psi(\mathbf{x})|^4 \Big]\psi , \label{Eq:GPE}
\end{equation}
where the kinetic and trap energies are contained within the single particle Hamiltonian,
\begin{equation}
H_0= \frac{-\hbar^2\nabla}{2m} + \frac{m}{2}\sum_j \omega_jx_j^2 .
\end{equation}
The contact and dipolar meanfield interactions are described, respectively, by
\begin{equation}
V_{\rm MF} = g|\psi(\mathbf{x})|^2 + \int d^3\mathbf{x}^\prime V_{\rm D}(\mathbf{x}-\mathbf{x}^\prime)|\psi(\mathbf{x}^\prime)|^2,
\end{equation}
where $g=4\pi\hbar^2a_s/m$, for s-wave scattering length $a_s$, mass $m$, and
\begin{equation}
V_{\rm D}(\mathbf{r})=\frac{\mu_0\mu^2}{4\pi r^3}(1-3\cos^2\theta) ,
\end{equation}
with $\mu$ being the magnetic dipole moment, and $\theta$ is the angle between $\mathbf{r}$ and the polarization direction $z$.
Meanfield collapse may be halted by the stabilizing effects of quantum fluctuations which, to leading order, can be described by the dipolar Lee-Huang-Yang correction \cite{Lima2011}
\begin{equation}
V_{\rm QF}(\mathbf{x}) = (32g|\psi(\mathbf{x})|^3/3)\sqrt{a_s^3/\pi}(1-3a_{dd}^2/2a_{\rm s}^2) .
\end{equation}
Stripe lifetimes are limited by 3-body losses characterized by the coefficient for a condensate $L_3$, which we measure experimentally.

We evolve the generalized Gross-Pitaevskii equation (1) using a $4$th-order Runge-Kutta method on a $256\times256\times64$ grid.
Since the dipole-dipole interactions are evaluated in Fourier space, a long-range cutoff is employed to prevent artificial interactions between Fourier copies, see \cite{Ronen2006}.

\subsubsection{Initial state preparation}\label{Sec:ISP}

Our initial states are created by first performing imaginary time evolution to find the ground state solution $\psi_0$ with $a_s=110a_0$ and $N=4\times10^4$. This is then randomly seeded with quantum and thermal fluctuations according to the truncated Wigner prescription \cite{Blakie2008}. In practice, this means adding the single particle modes $\phi_j$ according to
\begin{equation}
\psi(\mathbf{x},t=0)=\psi_0(\mathbf{x}) + \sum_j \beta_j \phi_j(\mathbf{x}) ,
\end{equation}
where $\beta_j$ are complex Guassian random variables obeying
\begin{equation}
\langle | \beta_j|^2 \rangle = \frac{1}{e^{\epsilon_n/k_BT}-1} + \frac{1}{2} .
\end{equation}
In our simulations we take the temperature to be $T=10$ nK which adds around 100 thermal atoms, consistent with the undetectable thermal fraction in our experiments. Although small, this initial noise plays an important role for seeding the instabilities that occur as the scattering length is ramped downwards. As in the experiments, we ramp $a_s$ to its final value over $30$ ms.
For a sufficiently small final $a_s$, we observe in our numerics the growth of a modulational instability, which for $a_s=94a_0$ has a wavelength of $\bar{k}_x = 1.6 \mu$m$^{-1}$, in agreement with the theory of roton instability discussed in Sec.~\ref{Sec:Ana}.

\subsubsection{Stripe contrast calculations}

\begin{figure}
\includegraphics[width=\columnwidth]{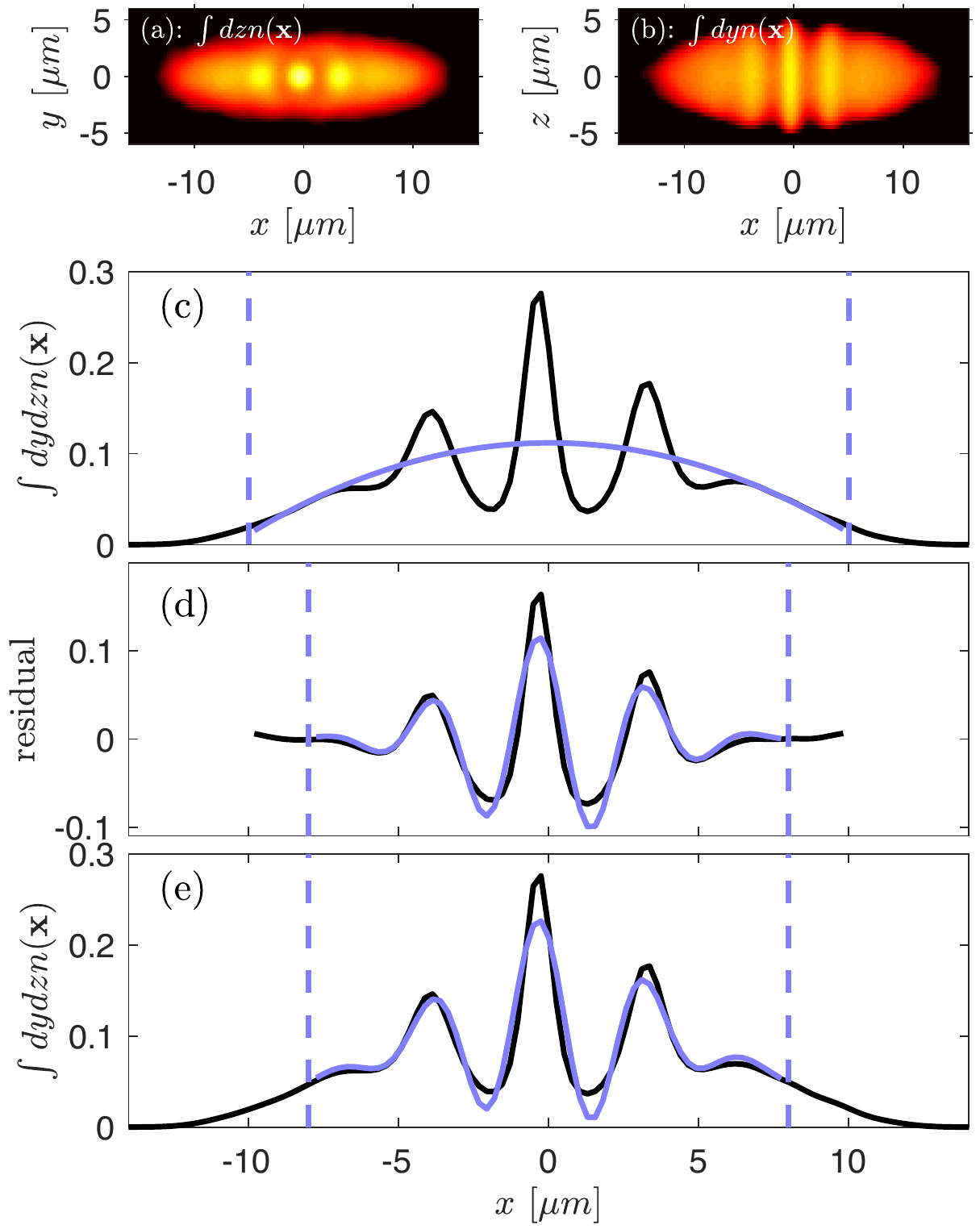}
\caption{(a) Column density integrated along $z$ on a logarithmic scale. (b) Similar column density but integrated along $y$. (c) Doubly integrated 1D density $n_{\rm 1D}(x)$ (black) and Thomas-Fermi fit $n_{\rm 1D}^{\rm TF}(x)$ (blue/gray). (d) Residual density $n_{\rm res} = n_{\rm 1D}(x)-n_{\rm 1D}^{\rm TF}(x)$ (black) and fitted function $n_{\rm res}^{\rm fit}(x)$ (blue/gray). (e) 1D density $n_{\rm 1D}(x)$ (black) compared with the combined fit $n_{\rm 1D}^{\rm fit} = n_{\rm 1D}^{\rm TF}(x) + n_{\rm res}^{\rm fit}(x)$ (blue/gray). The vertical dashed lines indicate the respective fit regions. This simulation snapshot is for $t=62$ ms and $a_s/a_0=94$ for one of the simulation runs presented in the main text.}  \label{StripeFitFig}
\end{figure}

In order to characterize the strength of the stripe modulations we perform fits to the doubly integrated 1D density $n_{\rm 1D}(x) = \int dy dz n(\mathbf{x})$, plotted as the black curve in Fig.~\ref{StripeFitFig} (c). The first is a Thomas-Fermi fit to capture the overall density profile,
\begin{equation}
n_{\rm 1D}^{\rm TF}(x) = a_1 - b_1 (x-c_1)^2 , \label{Eq:nTF}
\end{equation}
plotted as the blue/gray curve. Next we subtract the fit to obtain the residual density,
$n_{\rm res}(x) = n_{1D}(x) - n_{\rm 1D}^{\rm TF}(x)$ [the black curve in Fig.~\ref{StripeFitFig} (d)].
We then fit the residual using the function (plotted as the blue/gray curve)
\begin{equation}
n_{\rm res}^{\rm fit}(x) = a_2\sin (b_2x-c_2)e^{-d_2(x-e_2)^2} .
\end{equation}
Finally, we define the stripe contrast ($\mathcal{C}$) as the ratio of the fit amplitudes, i.e.,
\begin{equation}
{\rm \mathcal{C}} = \frac{a_2}{a_1} .
\end{equation}
For perspective, the combined fit $n_{\rm 1D}^{\rm fit}(x)=n_{\rm 1D}^{\rm TF}(x) + n_{\rm res}^{\rm fit}(x)$ (blue/gray curve) is compared against $n_{\rm 1D}(x)$ (black curve) in Fig.~\ref{StripeFitFig} (e).

\subsubsection{Phase incoherence calculations}

\begin{figure}
\includegraphics[width=\columnwidth]{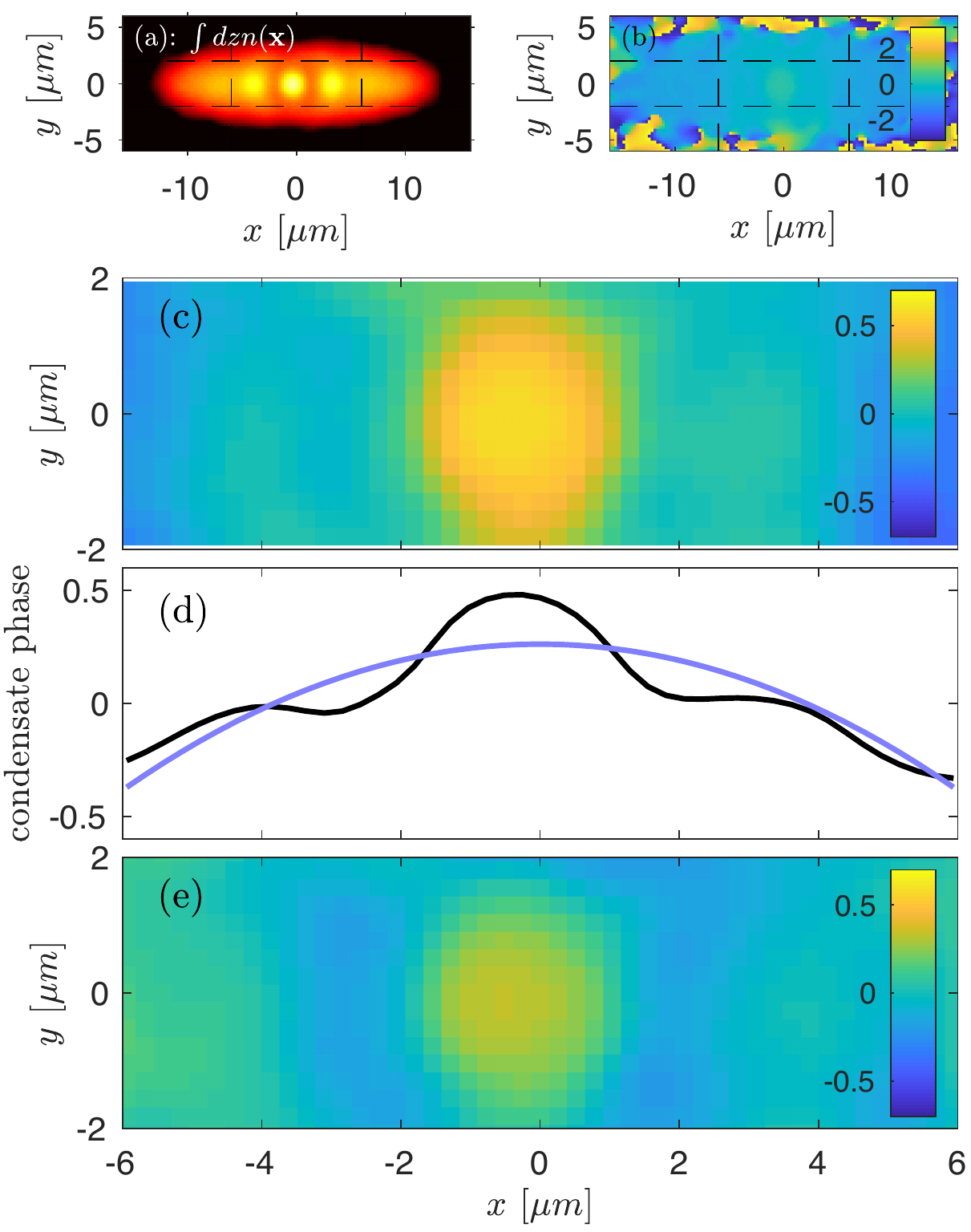}
\caption{(a) Column density integrated along $z$, on a logarithmic scale. (b) Wavefunction phase in the $z=0$ plane. (c) Zoom to central region of condensate phase in $z=0$ plane. (d) Condensate phase in central region but averaged along $y$ (black), and breathing mode fit (blue/gray). (e) Zoom of condensate phase residual $\alpha$ in the central region, obtained by subtracting the breathing mode. This simulation snapshot is for $t=62$ ms and $a_s/a_0=94$ for one of the simulation runs presented in the main text.}  \label{PhFitFig}
\end{figure}

Here we outline our approach for numerically quantifying the degree of phase coherence across the stripes. The first step is to calculate the wavefunction phase in the $z=0$ plane as shown in Fig.~\ref{PhFitFig} (b). Since we are primarily interested in the central region containing the stripes, Fig.~\ref{PhFitFig} (c) presents a zoom of the (recentered) phase in the region defined by $-6\leq x\leq 6~\mu m$ and $-2\leq y \leq 2~\mu m$ [indicated by the dashed lines in Fig.~\ref{PhFitFig} (b)]. An axial breathing excitation is clearly visible in the phase and, in order to remove this, we average the phase in this region along $y$ to obtain the 1D phase profile shown in Fig.~\ref{PhFitFig} (d) as the black curve. This is then fitted with a quadratic function of the same form as Eq.~(\ref{Eq:nTF}), shown as the blue/gray curve. By subtracting this fitted function we obtain a 2D phase residual $\alpha(x,y)$, see Fig.~\ref{PhFitFig} (e). Finally, the phase incoherence is defined by
\begin{equation}
\alpha^I = \frac{\int^\tau dx dy n(x,y) |\alpha(x,y) - \langle \alpha(x,y) \rangle |}{\int^\tau n(x,y) dx dy} ,
\end{equation}
where $\langle\alpha(x,y)\rangle$ is the average phase for a given realization, and $\tau$ is defined as the region $-6\leq x \leq 6~\mu m$. A value of $\alpha^I=0$ implies perfect global phase coherence, whereas $\alpha^I=\pi/2$ indicates incoherence.

\subsection{Simulation results without 3-body losses}

\begin{figure}
\includegraphics[width=\columnwidth]{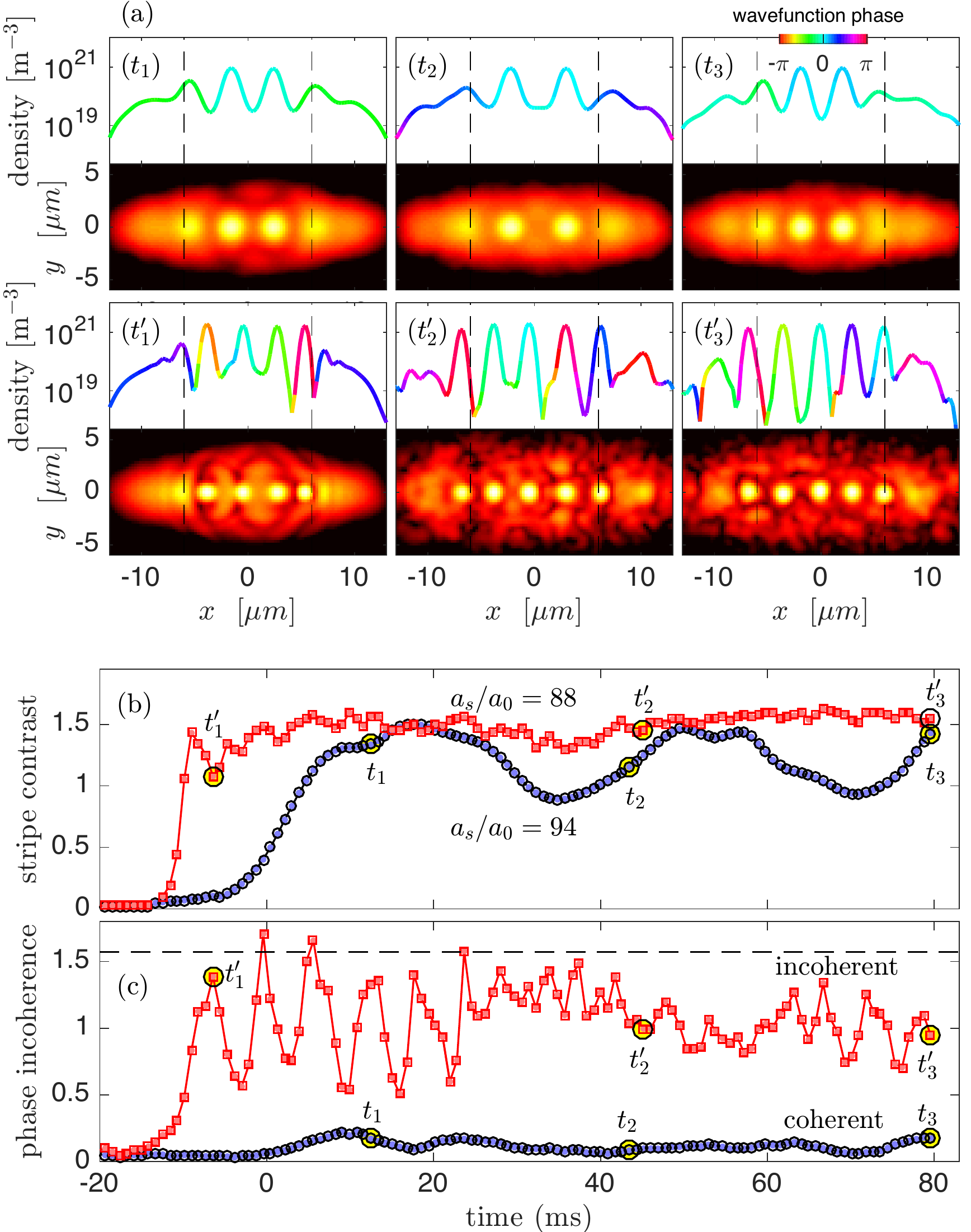}
\caption{Two simulations of stripe formation and evolution in the absence of losses ($L_3=0$). (a) Snapshots after a ramp to $a_s/a_0=94$ are shown at times $(t_1,t_2,t_3)=(12.5,43.5,79.6)$ ms, while incoherent stripes can be seen after a ramp to $a_s/a_0=88$ at $(t_1^\prime,t_2^\prime,t_3^\prime)=(-6.4,45.2,79.6)$ ms. Density cuts $n(x,0,0)$ -- with color representing wavefunction phase -- and column densities $\int dz n(x,y,z)$ are shown.  (b) Stripe contrast for $a_s/a_0=94$ (blue circles) and $a_s/a_0=88$ (red squares). (c) Phase incoherence, $\alpha^I=\int^\tau dxdy n|\alpha - \langle \alpha \rangle |/\int^\tau dxdy n $, where $\alpha(x,y,0)$ is the wavefunction phase, $\langle\alpha\rangle$ is its average for a given microstate, and $\tau$ is the region indicated by dashed lines in the upper panels. The limit $\alpha^I=0$
indicates global phase coherence while $\alpha^I=\pi/2$ signals incoherence.}  \label{K0Fig}
\end{figure}

Here we investigate the role of 3-body losses by performing simulations in their absence, i.e.~we set $K_3=0$. Figure \ref{K0Fig} shows the results of two simulations for the same two scattering length ramps considered for the simulations in the main text.
The one with the ramp to $a_s/a_0=94$ [blue circles in Figs.~\ref{K0Fig} (b)(c)] is very similar to the ones we presented for the case with $K_3\neq0$. However, instead of the eventual stripe decay that we saw there, here they persist throughout the simulation and show no sign of weakening. This provides further evidence that the supersolid properties observed in our experiments are very robust, and their eventual decay is due to 3-body loses. Note that even without losses, a long-lived axial breathing oscillation is triggered by the droplet formation, but this does not break the phase locking between the droplets.
In contrast, the simulation with the ramp to $a_s/a_0=88$ [red squares in Figs.~\ref{K0Fig} (b)(c)] is markedly different to the corresponding simulations in the main text with $K_3\neq0$. While there the stripe contrast and phase locking were both rapidly lost almost immediately after droplet formation, here the lossless simulation exhibits robust stripe contrast throughout the simulation, although the phase coherence between droplets is still rapidly lost. The reason for the loss of phase coherence can clearly be seen in Figs.~\ref{K0Fig} ($t_1^\prime-t_3^\prime$), where the high density droplets are separated by very low density regions.



\end{document}